\documentclass{article}

\usepackage[final]{neurips_2023}

\usepackage[utf8]{inputenc} 
\usepackage[T1]{fontenc}    
\usepackage{hyperref}       
\usepackage{url}            
\usepackage{booktabs}       
\usepackage{amsfonts}       
\usepackage{nicefrac}       
\usepackage{microtype}      
\usepackage{xcolor}         
\usepackage{caption}
\usepackage{subcaption}
\usepackage{graphicx} 
\usepackage{varwidth}
\usepackage{wrapfig}
\DeclareCaptionFormat{centerformat}{%
  \begin{varwidth}{\linewidth}%
    \centering
    #1#2#3%
  \end{varwidth}%
}
\usepackage{tikz}
\usepackage{multirow}
\usepackage{floatrow}
\usepackage[export]{adjustbox}
\usepackage{graphicx}

\title{Ultra-Resolution Cascaded Diffusion Model for Gigapixel Image Synthesis in Histopathology}

\usetikzlibrary{shapes.geometric, arrows, positioning}

\tikzstyle{blue} = [rectangle, rounded corners, minimum width=8em, minimum height=1.8em,text centered, draw=black, fill=blue!30]
\tikzstyle{yellow} = [rectangle, rounded corners, minimum width=8em, minimum height=1.8em,text centered, draw=black, fill=yellow!30]
\tikzstyle{green} = [rectangle, rounded corners, minimum width=8em, minimum height=1.8em,text centered, draw=black, fill=green!30]
\tikzstyle{arrow} = [thick,->,>=stealth]

\hypersetup{
    colorlinks=true,
    urlcolor=blue,
    linkcolor=black,  
    citecolor=black,
}

\author{%
  Sarah~Cechnicka\\
  ICL, 
  London, UK \\
  \And
  Hadrien~Reynaud\\
  ICL,
  London, UK 
  \And
  James~Ball\\
  ICL,
  London, UK 
  \And
  Naomi~Simmonds\\
  NHS Trust, 
  London, UK \\
  \And
  Catherine~Horsfield\\
  NHS Trust,
  London, UK \\
  \And
  Andrew~Smith\\
  NHS Trust, 
  London, UK \\
  \And
  Candice~Roufosse\\
  ICL,
  London, UK \\
  \And
  Bernhard~Kainz\\
  ICL,
  London, UK \\
}

\usepackage[acronym]{glossaries}
\newacronym{wsi}{WSI}{Whole Slide Image}
\newacronym{cdm}{CDM}{Cascaded Diffusion Model}
\newacronym{urcdm}{URCDM}{Ultra-Resolution Cascaded Diffusion Model}
\newacronym{lrdm}{LRDM}{Lower Resolution Diffusion Model}
\newacronym{mae}{MAE}{Mean Absolute Error}
\begin{document}
\maketitle

\begin{abstract}
Diagnoses from histopathology images rely on information from both high and low resolutions of Whole Slide Images. Ultra-Resolution Cascaded Diffusion Models (URCDMs) allow for the synthesis of high-resolution images that are realistic at all magnification levels, focusing not only on fidelity but also on long-distance spatial coherency. Our model beats existing methods, improving the pFID-50k~\cite{anyresgan} score by 110.63 to 39.52 pFID-50k. Additionally, a human expert evaluation study was performed, reaching a weighted \gls{mae} of 0.11 for the Lower Resolution Diffusion Models and a weighted \gls{mae} of 0.22 for the URCDM.
\end{abstract}

\section{Introduction}
In healthcare, access to data is often restricted due to privacy concerns, making it difficult to develop robust models. Diffusion models have proven particularly valuable in this context, allowing for the generation of synthetic data while preserving patient confidentiality. In histopathology, where staining differences hinder generalizability, diffusion models achieve a more dependable performance across diverse data sources~\cite{Cechnicka2023}. 
Most research focuses on the use of generative approaches at patch level, where these techniques excel at generating high-fidelity, localized details~\cite{ganreview,macenko,Federated}. However, significant challenges arise when transitioning from patch-level analysis to \glspl{wsi}, such as memory constraints, long sampling times, and a lack of training data. Medical professionals, however, often rely on both low and high-magnification views to make critical decisions. As such, generation at \gls{wsi} level is important. By increasing the number of data samples available, it enables the use of more complex downstream algorithms, which operate on the entire image at different scales, e.g., You Only Look Twice~\cite{yolt}.  This transition introduces complexities related to preserving contextual information and ensuring seamless integration between different magnification levels.
Recent applications of cascaded diffusion models~\cite{imagen} have improved multi-scale information integration in medical imaging, pushing the boundaries of diagnostic accuracy and patient care ~\cite{reynaud2023featureconditioned}. These diffusion models are chained together, first generating a small image, which gets repeatedly upsampled, once by each model in the cascade. This process allows for smaller models and parallel training, making these large-scale experiments much more tractable in time and compute resources.
\section{Method}

\glspl{cdm} work by first generating a low-resolution image $I_0$ with a base model $C_{\phi_0}$ from Gaussian noise. $I_0$ is subsequently used as a conditional input for a second diffusion model $C_{\phi_1}$\cite{ondistillationofguideddiffusionmodels}, which generates an image of higher resolution $I_1$, that is similar to $I_0$. Generally, the nth super-resolution stage will condition on the lower-resolution image $I_{n-1}$ generated in the previous stage. We can write
$
C_{\phi_n}\left(\sigma, I_{n-1}\right)=F_{\phi_n}\left(c_{\text {noise }}(\sigma), I_{n-1}\right), C_{\phi_0}\left(\sigma\right)=F_{\phi_0}\left(c_{\text {noise }}(\sigma)\right).
$



\begin{figure}[h]
    \centering
    \includegraphics[width=\linewidth]{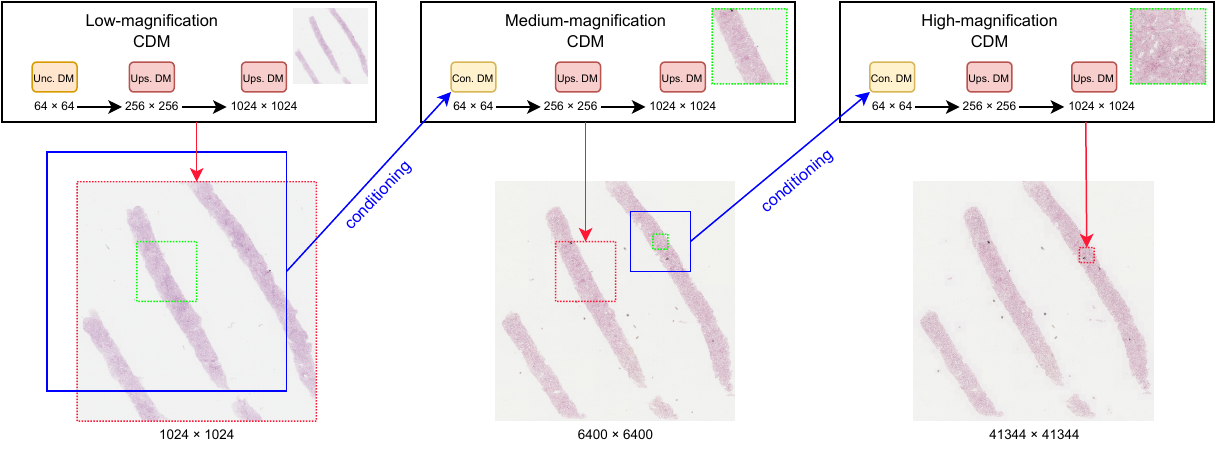}
    \caption{Detailed overview of the \gls{urcdm} image generation process. The medium and high-magnification \glspl{cdm} are sampled many times, and the patches generated are stitched together; one sample shown as an example. A blue outline indicates the lower-magnification conditioning image, meant to teach the context for the new generation process. A green outline indicates the resultant patch that will be `zoomed in' on and generated by a baseline \gls{cdm}. Red lines indicate the output of each magnified image. Not to scale.}
    \label{fig:urcdm-gen-process}
\end{figure}

\glspl{urcdm} use a similar principle, but extend it further. First, a base \gls{wsi} is generated at low resolution, and upsampled multiple times using the described \gls{cdm} approach, to reach a low-magnification resolution. Then, the low-magnification \gls{wsi} is split into patches with a certain degree of overlap. Those patches are used for conditioning a new base diffusion model, trained to generate smaller patches at the centre of each low-magnification patch through further upsampling, by using another \gls{cdm}. These higher-resolution patches are stitched back together to produce a seamless medium-magnification image. The same process is repeated over, through yet another \gls{cdm} to reach the final high-resolution magnification image, at gigapixel scale. Through this setup, long-distance coherence in enforced, and all models keep a tractable size. This process is illustrated in Figure~\ref{fig:urcdm-gen-process}.

In the \gls{urcdm}, there are 9 diffusion models in total, for three separate \glspl{cdm}. 
All models are trained independently and in parallel on 9 separate Nvidia A100 GPUs.
The models' architecture and training are heavily based on Imagen~\cite{imagen} using the \verb|imagen-pytorch|~\cite{imagenpytorch} library. Each \gls{cdm} targets a different magnification of the overall image. 
At inference time, a low quality \gls{wsi} is generated with the use of the low-resolution \gls{cdm}, the output of which has the size $1024 \times 1024$. Overlapping patches of this generated low-resolution image are then used for conditioning of the second \gls{cdm}, by giving it the spatial context needed for generation. The second model generates images of size $1024 \times 1024$ for the centre or each conditioning patch and after stitching them together, and accounting for overlaps, the resolution of the \gls{wsi} increases to $6400 \times 6400$. This process is repeated for the final high-resolution model, yielding a final synthetic \gls{wsi} of $41344 \times 41344$ pixels. The size of the medium-magnification \gls{cdm} was arbitrarily chosen as being approximately halfway between low-resolution and high-resolution magnifications. The \gls{urcdm} is not restricted to these resolutions, which can trivially be changed to suit other datasets.


Gradient clipping was implemented (set to 1) as well as v-parametrisation, to avoid lower quality fine details, blurring or heavy distortion of lower magnification images when `zooming in'. Outpainting is used to smoothly merge generated patches together with minimal seams. The resulting dependency between patches is especially important when batch-processing images or sampling on multiple GPUs, as it determines when patches can be generated in parallel, which ensures reasonable sampling speeds. 12.5\% of overlap (both in the vertical and horizontal directions) was chosen to minimize patching artefacts while reasonably decreasing the total number of patches generated. High-magnification patches of the whole slide image that are mostly white are ignored in both training and sampling. Instead, all white patches are replaced with an upscaled version of the medium-magnification image. All images were cropped or padded to $40,000 \times 40,000$ pixels.
\section{Results}
Two main measurements of success are established. Firstly, ultra-resolution images must look realistic at multiple scales both when ‘zoomed-out’ and when ‘zoomed-in'. This is done by calculating FID scores and comparing them against two baselines. Lower-resolution patches, are evaluated against StyleGAN3~\cite{stylegan3}. 
Unconditional \gls{lrdm} with a final FID-10k score of 10.35 significantly outperform GANs which reach an FID-10k score of only 38.62. This is likely due to GAN's often causing unnatural symmetries to appear in the images, including ringing artefacts. Second, long-distance spatial coherency is considered. The realism of high-resolution images is compared against baseline unconditional diffusion models generating high-resolution images using outpainting. The pFID of the images generated by the \gls{urcdm} is 39.52 much lower than those generated using outpainting which is  150.15. Whilst when zoomed in, fine details of the images generated using outpainting are of higher quality when analysed qualitatively, the lack of spatial coherency leads to much poorer pFID scores. 

\begin{table}[t]
\begin{tabular}{p{1em}p{5.5em}p{1.5em}p{1.5em}p{2.5em}p{4em} p{1em}p{1.5em}p{1.5em}p{2.5em}p{4em}}
\toprule
& User &TP & FP & $p$ &$|p-0.5|$  & & TP & FP & $p$ &$|p-0.5|$   \\ \midrule
 \multirow{5}{*}{\rotatebox{90}{\gls{lrdm}}}
& Pathologist 1 & 250  & 179    & 0.4172    & 0.0823 & 
\multirow{5}{*}{\rotatebox{90}{\gls{urcdm}}} 
&  61   & 66    & 0.5197    & 0.0197  \\  
&Pathologist 2 & 106    & 145   & 0.5777  & 0.0777 &  
& 152  & 6   & 0.0380  & 0.4620 \\ 
&Pathologist 3 & 29     & 99   & 0.7734   & 0.2734 &  
& 28   & 33   & 0.5410   & 0.0410  \\ 
&Pathologist 4 &-&-&-&- & 
&  47   & 10   & 0.1754   & 0.3246  \\ 
&Non-expert    & 110  & 162    & 0.5956      & 0.0956 & 
& 29  & 21    & 0.4200   & 0.0800\\ \midrule
&Total & 495 & 585 & 0.5417 & 0.1074 \newline (w-\gls{mae})& 
&317 & 136 & 0.3002 & 0.2219 \newline (w-\gls{mae}) \\ 
\bottomrule
\end{tabular}
\caption{Results of human evaluation of the realism of synthetic images generated by an unconditional \gls{lrdm} as well as random crops of synthetic images generated by the \gls{urcdm}.TP stands for true positives, FP for false positives and $p$ denotes the proportion of incorrectly classified samples. W-\gls{mae} stands for the weighted \gls{mae}.\label{tab:human-eval}}
\end{table}

\begin{wrapfigure}{r}{0.40\textwidth}
  \begin{center}
  \vspace{-20pt}
\includegraphics[angle=90,width=1.0\linewidth]{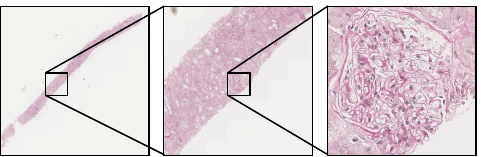}
  \end{center}
  \caption{Sample of an ultra-resolution whole slide image generated using a URCDM. Full-scale ultra-resolution image ($40000 \times 40000$ pixels), partially zoomed-in crops of that image ($6500 \times 6500$ pixels) and highly zoomed-in crops of that image ($1024 \times 1024$ pixels).  \vspace{-10pt}}\label{fig:urcdm-kidney}
\end{wrapfigure}
Additionally, images of various resolutions are evaluated by pathologists on their realism. We can see from Table \ref{tab:human-eval} that the proportion of incorrectly classified samples $p$ varies quite a lot between users, with some having a strong preference for real images, and others having a strong preference for fake images. This balances out with the total proportion being close to 0.5 for \gls{lrdm}. Expert pathologist evaluation was also carried out on ultra-resolution images. Results here are quite inconsistent among pathologists, with some identifying fake images more reliably than others. Due to the limited number of ultra-resolution images generated, they are more difficult to evaluate. To combat this, pathologists were shown a subset of the crops taken to calculate the pFID. 
Figure \ref{fig:urcdm-kidney} shows a sample image generated by the \gls{urcdm}. The texture of the tissue is realistically wispy in regions of the kidney further from the cortex, like in a real \gls{wsi}. Additionally, structures that are only visible at high magnification are of high quality, like the glomerulus and tubules in the final image of Figure \ref{fig:urcdm-kidney}. In general, at high magnification levels, there is a lot of diversity between different regions, whilst maintaining a good level of quality.
Despite successful results, the fine details in the high-resolution image of Figure \ref{fig:urcdm-kidney} are visibly poorer than those in the real \glspl{wsi}. This is likely due to \glspl{urcdm} having to maintain consistency through all nine stages. Base U-Nets are required to consistently and accurately `zoom in' the lower magnification image, and failure to do so will result in neighbouring patches looking very different.

\section{Conclusion}
\glspl{urcdm} are a novel and promising way of generating images with more than 1, 000, 000, 000 pixels using diffusion models. Images generated by \glspl{urcdm} are spatially coherent over long distances and images are plausible at different scales.
Fine details remain clear with URCDMs, which produce more coherent images than the outpainting method, crucial for ultra-resolution imagery applications that use various image scales~\cite{yolt}.
Future work will focus on computation efficiency, use of the same \gls{cdm} for all magnifications, and multi-modal learning.
\section{Potential Negative Societal Impact}
The development and deployment of the Ultra-Resolution Cascaded Diffusion Model (URCDM) for Gigapixel Image Synthesis in Histopathology has several potential societal and clinical implications.

One of the most significant impacts is in the realm of medical training and research. The ability to generate high-resolution synthetic histopathology images can revolutionize the way medical students and pathologists are trained. Instead of being limited to a finite number of samples, these professionals can access an almost unlimited array of images for study and diagnosis practice. This, in turn, could lead to more competent professionals and more accurate diagnoses.

Another implication is the potential to democratize access to medical education resources. In many parts of the world, medical training facilities lack the necessary resources to provide students with access to a wide variety of histopathology samples. With tools like \gls{urcdm}, these institutions could have access to a vast library of synthetic images, reducing the gap between well-funded and under-resourced medical schools.

From a research perspective, having a tool that can generate a vast number of high-resolution images can also expedite the discovery process. For instance, when researching rare diseases or conditions, scientists and medical professionals may not always have access to a large number of samples. In such cases, the \gls{urcdm} could be invaluable.

However, with such a powerful tool, there are also ethical considerations to address. There is a risk that the generated images could be misused or misrepresented in research or other professional settings. Moreover, while the synthetic images may resemble real samples, they are not. As such, relying solely on them without validation against actual samples could lead to erroneous conclusions. Striking a harmonious balance between pioneering innovation and the imposition of rigorous ethical standards becomes imperative, necessitating the establishment of comprehensive regulatory frameworks. 

Additionally, what demands consideration is the intrinsic aptitude of generative models to acquire and, potentially, recreate specific patient-related information. While these models are crafted to produce synthetic, de-identified patient images, the depth of their learning could inadvertently lead to the generation of images bearing striking similarities to the original dataset, potentially exposing confidential patient data. This becomes an even larger problem with gigapixel images, as limited training data exists, and overfitting to the real images becomes a problem. This prospect engenders supplementary ethical inquiries regarding the extent of information assimilated and retained by such models, as well as the inadvertent revelation of identifiable patient data. 

\noindent\textbf{Acknowledgements:} 
This work was supported the UKRI Centre for Doctoral Training in Artificial Intelligence for Healthcare (EP/S023283/1). Dr.Roufosse is supported by the National Institute for Health Research (NIHR) Biomedical Research Centre based at Imperial College Healthcare NHS Trust and Imperial College London. The views expressed are those of the authors and not necessarily those of the NHS, the NIHR or the Department of Health. 

\bibliographystyle{plain}
\bibliography{bibliography.bib}

\end{document}